\documentclass[aps,prl,twocolumn,showpacs,superscriptaddress,amsmath,amssymb,10pt]{revtex4-1}

\usepackage{graphicx}
\DeclareGraphicsExtensions{.pdf,.png,.jpg}

\usepackage[normalem]{ulem}
\usepackage{xcolor}

\begin{document}

\title{\Large{Self-injection by trapping of plasma $e^-$ oscillating in rising density gradient at the vacuum-plasma interface}}

\author{Aakash A. Sahai}
\email{aakash.sahai@gmail.com}
\author{Thomas C. Katsouleas}
\affiliation{Dept of Electrical Engineering, Duke University, Durham, NC, 27708 USA}
\author{Patric Muggli}
\affiliation{Max-Planck-Institut fur Physik, 80805 Munchen, Germany}

\begin{abstract}
We model the trapping of plasma $e^-$ within the density structures excited by a propagating energy source ($\beta_{S}\simeq1$) in a rising plasma density gradient. Rising density gradient leads to spatially contiguous coupled up-chirped plasmons ($d{\omega^2_{pe}(x)}/{dx}>0$). Therefore phase mixing between plasmons can lead to trapping until the plasmon field is high enough such that $e^-$ trajectories returning towards a longer wavelength see a trapping potential. Rising plasma density gradients are ubiquitous for confining the plasma within sources at the vacuum-plasma interfaces. Therefore trapping of plasma-$e^-$ in a rising ramp is important for acceleration diagnostics and to understand the energy dissipation from the excited plasmon train \cite{LTE-2013}. Down-ramp in density \cite{density-transition-2001} has been used for plasma-$e^-$ trapping within the first bucket behind the driver. Here, in rising density gradient the trapping does not occur in the first plasmon bucket but in subsequent plasmon buckets behind the driver. Trapping reduces the Hamiltonian of each bucket where $e^-$ are trapped, so it is a wakefield-decay probe. Preliminary computational results for beam and laser-driven wakefield are shown.

\end{abstract}

\maketitle

\section{Introduction}

In this paper we show the self-trapping of plasma $e^-$ in a plasma acceleration structure due to phase-mixing in an up-ramp density inhomogeneity at the vacuum-plasma interface. Rising density inhomogeneities at the vacuum-plasma interface are universal features of plasma sources. Hence, it is critical to study the plasma processes in a density gradient. Plasma acceleration structures are excited by coherent motion of the plasma $e^-$ driven by resonant energy sources in the plasma. Plasma $e^-$ typically oscillate within the potential well of the background ions spatially limited to a single bucket and do not co-propagate with the driver. Trapping is a mechanism of self-injection of plasma $e^-$ into the plasma acceleration structure resulting in the transfer of energy (and acceleration) to the  trapped electron beam. The trapped $e^-$ propagate across many plasma-wavelenghts locked to the crest of the wakefield. Therefore beam loading of the acceleration structure potential leads to dissipation from the wakefields\cite{LTE-2013}. The highest field amplitude plasmon-bucket is just behind the driver. The buckets subsequent to the first have smaller fields as they dissipate energy to the surrounding plasma and also to the plasma ions\cite{LTE-2013}. Most optimized injection schemes like external injection, self-injection due to non-linear plasmon oscillations and returning trajectory crossing, ionization injection, down-ramp injection\cite{density-transition-2001}, colliding-pulse injection etc. inject into the first wakefield bucket to accelerate at the peak gradient.

\begin{figure}
	\begin{center}
   	\includegraphics[width=3.25in]{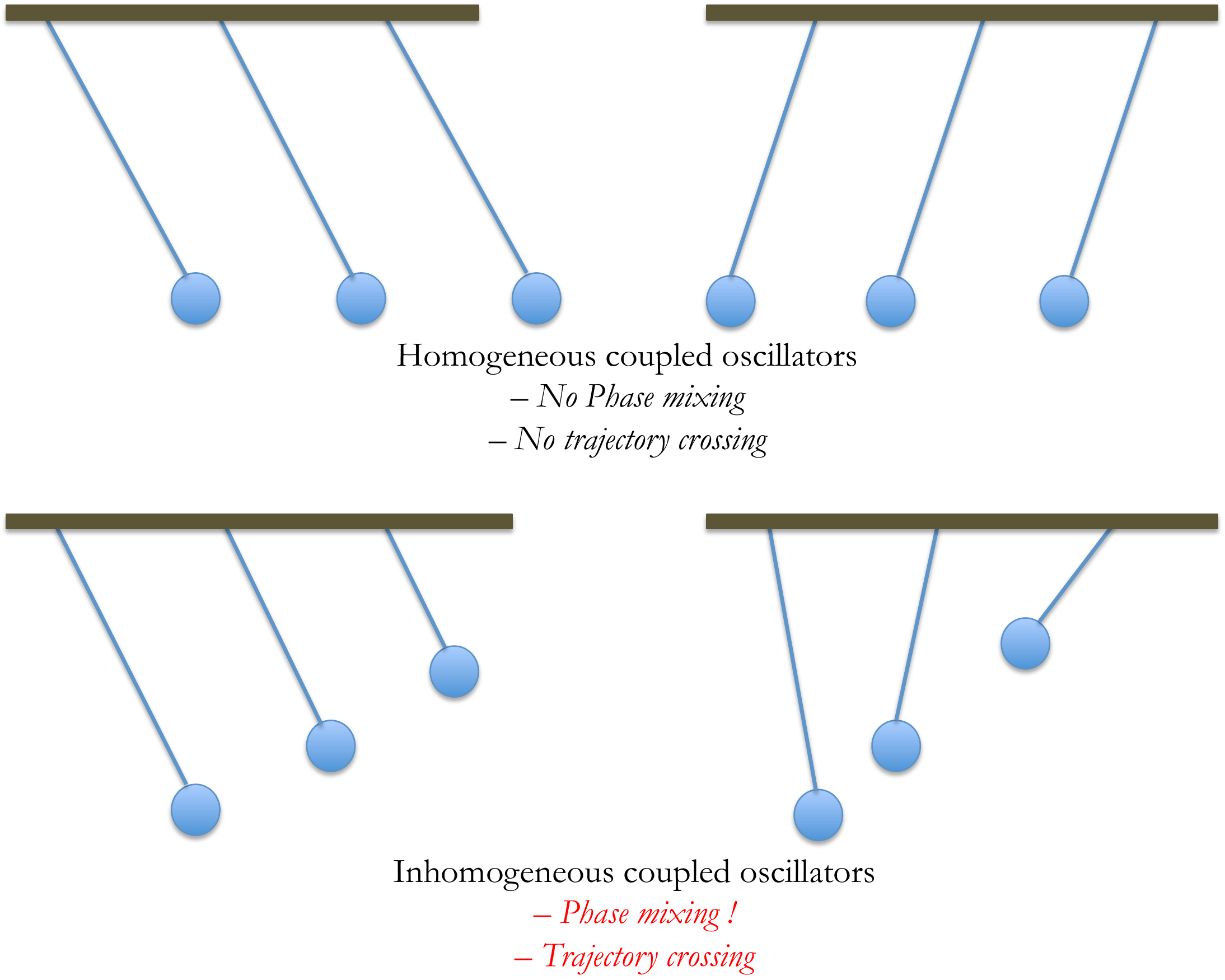}
	\end{center}
\caption{ {\it Phase mixing of inhomogeneous coupled oscillators with spatially increasing frequencies}. Individual plasmons excited in the wake of a driver in a homogeneous plasma are coupled together and undergo in-phase or synchronous-phase oscillations. However, in an inhomogeneous rising density plasma (common at the vacuum-plasma interface of plasma sources) the individual plasmons have increasing characteristic frequency. Hence there is phase-mixing and trajectory crossing.}
\label{fig:phase-mixing-cartoon}
\end{figure}

It is of relevance to study the injection of plasma $e^-$ in a rising density inhomogeneity. An uncontrolled injection of plasma $e^-$ in successive plasmon buckets can increase the energy spread of the accelerated beam. The termination of injection is critical to maintaining nearly equal acceleration lengths at peak field of the stacked trapped beams injected into multiple plasmon-buckets. Secondly, in beam-driven schemes where the forward longitudinal momenta is smaller in comparison to ultra-short laser pulses (where forward longitudinal ponderomotive force is higher), this mechanism can lead to injection of plasma $e^-$. The beam-driven plasma does not trap and self-inject because plasma electron longitudinal momenta are lower. These trapped particles may be observed as beam-lets at driving beam energy. And, an analysis of trapped beams can be used as a diagnostics of the beam-plasma interaction and also to study the state of the plasma behind the driver\cite{LTE-2013}.

\section{Phase-mixing of plasmons}

As the phase-mixing injection mechanism is a 1-D process it is described using 1-D model (seen in the on-axis trapping in Fig.\ref{fig:long-momentum-self-trapping}[a]-[c]). The 1-D momentum equation of longitudinal oscillations of a plasmon (along z) is $\frac{d^2}{d\tau^2}\left(\beta_{\phi} p_{ez} - \gamma_e \right) + \omega_{pe}^2(z) \beta^2_{\phi} \frac{ p_{ez} }{\beta_{\phi}\gamma_e - p_{ez} } = 0$, where $d\tau = d(z/\beta_{\phi} - t)$. By simplifying, $\beta^e_z=p_{ez}/\gamma_e$, $\beta_{\phi}\sim 1$ and $\mathcal{Z} = \sqrt{(1-\beta^e_z)/(1+\beta^e_z)}$, we have, $\frac{d^2}{d\tau^2}\mathcal{Z} + \omega_{pe}^2(\beta_{\phi}\tau + t) \frac{1}{2}(1/\mathcal{Z}^2 - 1)  = 0$. The equation is similar in form to the Hill's equation due to the presence of a function of dependent variable as the coefficient. The momentum solutions have phase-mixing (in Fig.\ref{fig:phase-mixing-cartoon}), as the solutions depend upon of plasmon oscillation frequencies and amplitudes ($k_{pe}(z)$ and $\omega_{pe}(z)$).

We can build a physical picture of trapping of plasma $e^-$ by phase mixing of trajectories. The up-ramped density inhomogeneity leads to frequency up-chirped spatial-train of coherently oscillating sheets. The perturbed infinitesimal plasma sheets have different characteristic plasma frequency due to different density, therefore coupled individual sheet oscillators undergo a phase-mixing (see Fig.1)\cite{dawson-phase-mixing-1959}. Following \cite{dawson-phase-mixing-1959}sec.IV-1D model, the displacement, $X$ from equilibrium, $x_0$ or the trajectory is $\frac{d^2X}{dt^2} = -\int_{x_0}^{x_0+X}\frac{4\pi e^2}{m_e} n_0(x) dx$. The time for mixing to start is $t_{mix}=\frac{\pi}{2(d\omega_{pe}/dx)X}$. The displacement from equilibrium, X depends upon the maximum longitudinal field, $\lambda_{pe}(E_{\parallel}/E_{WB})$. So, in 1-D model $t_{mix}\propto \frac{1}{E_{\parallel}/E_{WB}} \frac{1}{(d\omega_{pe}/dx)}$. The effect of phase-mixing between infinitesimal sheet leads to plasma $e^-$ encountering unbalanced fields that are not experienced during in-phase oscillations, this results in their disruption. The trajectory of first oscillation in the forward direction of propagation of the source do not cross. Because, the driven $e^-$ propagate in the same direction, $z_{forward}$. However, during the returning trajectories of the oscillations, $z_{returning}$, the faster oscillators (higher in density) encounter the potential well of the slower oscillators. Therefore the trapping of $e^-$ occurs in the second and subsquent buckets and not in the first bucket. Since, the peak fields are in the first bucket this trapping mechanism is not ideal for acceleration.

The disruption of the plasma oscillations also occurs when the $e^-$ trajectories are highly non-linear ($p_e>m_ev_{\phi}$) resulting in phase-mixing\cite{dawson-phase-mixing-1959} (for instance they cross in the back of the bubble). The plasma $e^-$ that go out-of-phase see unbalanced fields in the wave-frame and get trapped in the accelerating phase of the fields. Therefore in laser excited non-linear plasmons the trapping also occurs in the first bucket.

\section{Simulations - Laser and Beam driven gradient at the Vacuum-plasma interface}

\begin{figure}
	\begin{center}
   	\includegraphics[width=3.25in]{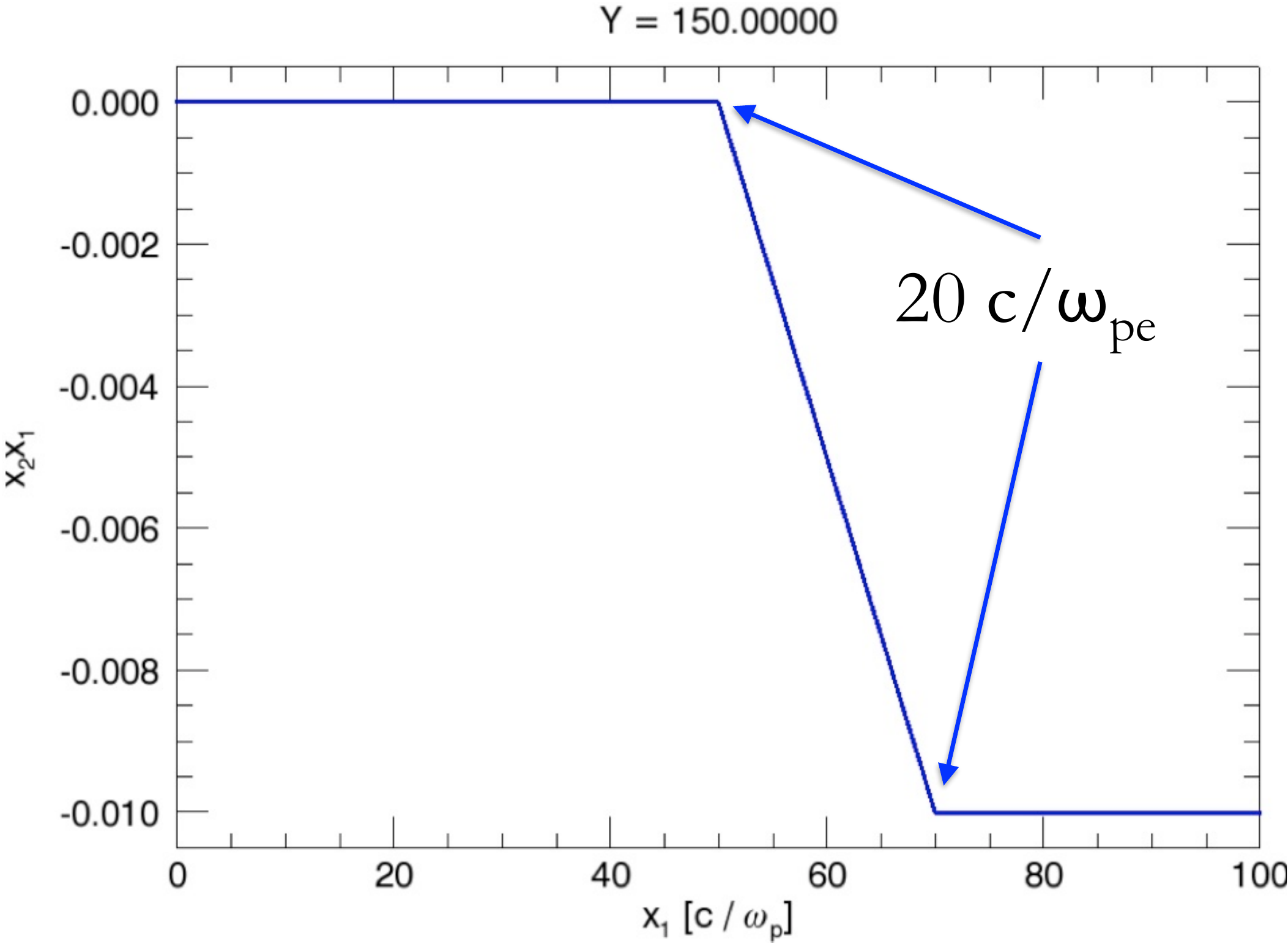}
	\end{center}
\caption{ {\it Rising plasma density gradient at the vacuum-plasma interface}. The $e^-$ density (negative) rises from 0 to $0.01n_c$ between 50 and 70$\frac{c}{\omega_{pe}}$. The homogeneous plasma density is $0.01n_c$.}
\label{fig:vacuum-plasma-gradient}
\end{figure}

\begin{figure*}
	\begin{center}
   	\includegraphics[width=6.5in]{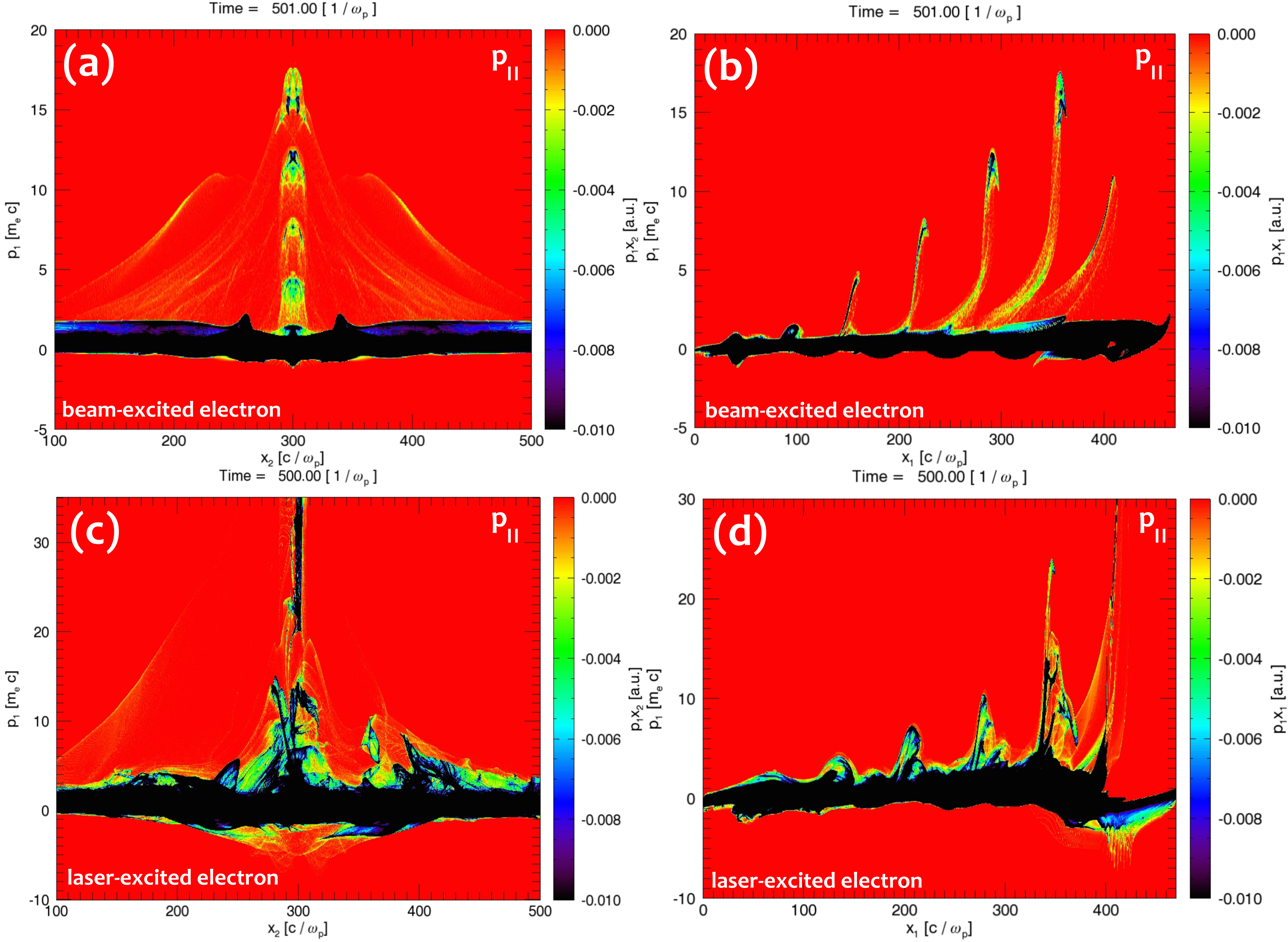}
	\end{center}
\caption{ {\it Longitudinal momentum ($p_{\parallel}=p_1$) of plasma $e^-$ in beam and laser driven wake-fields in 2-D PIC simulations}. The plasma $e^-$ longitudinal momentum phase-space is shown at $t\sim500\frac{1}{\omega_{pe}}$. Beam-driven phase-spaces are in (a) with transverse dimension ($p_1x_2$) and (b) with longitudinal ($p_1x_1$) dimension. Corresponding Laser-driven phase-space are in (c) and (d). The $e^-$ trapped in first laser-driven bucket gain a peak momentum $\gamma_{\parallel}\beta_{\parallel} > 30$, whereas in both the beam and laser case, the second bucket $e^-$ only gains, $\gamma_{\parallel}\beta_{\parallel} \sim 20$. Also, phase-mixing injection occurs only on the axis.}
\label{fig:long-momentum-self-trapping}
\end{figure*}

\begin{figure*}
	\begin{center}
   	\includegraphics[width=6.5in]{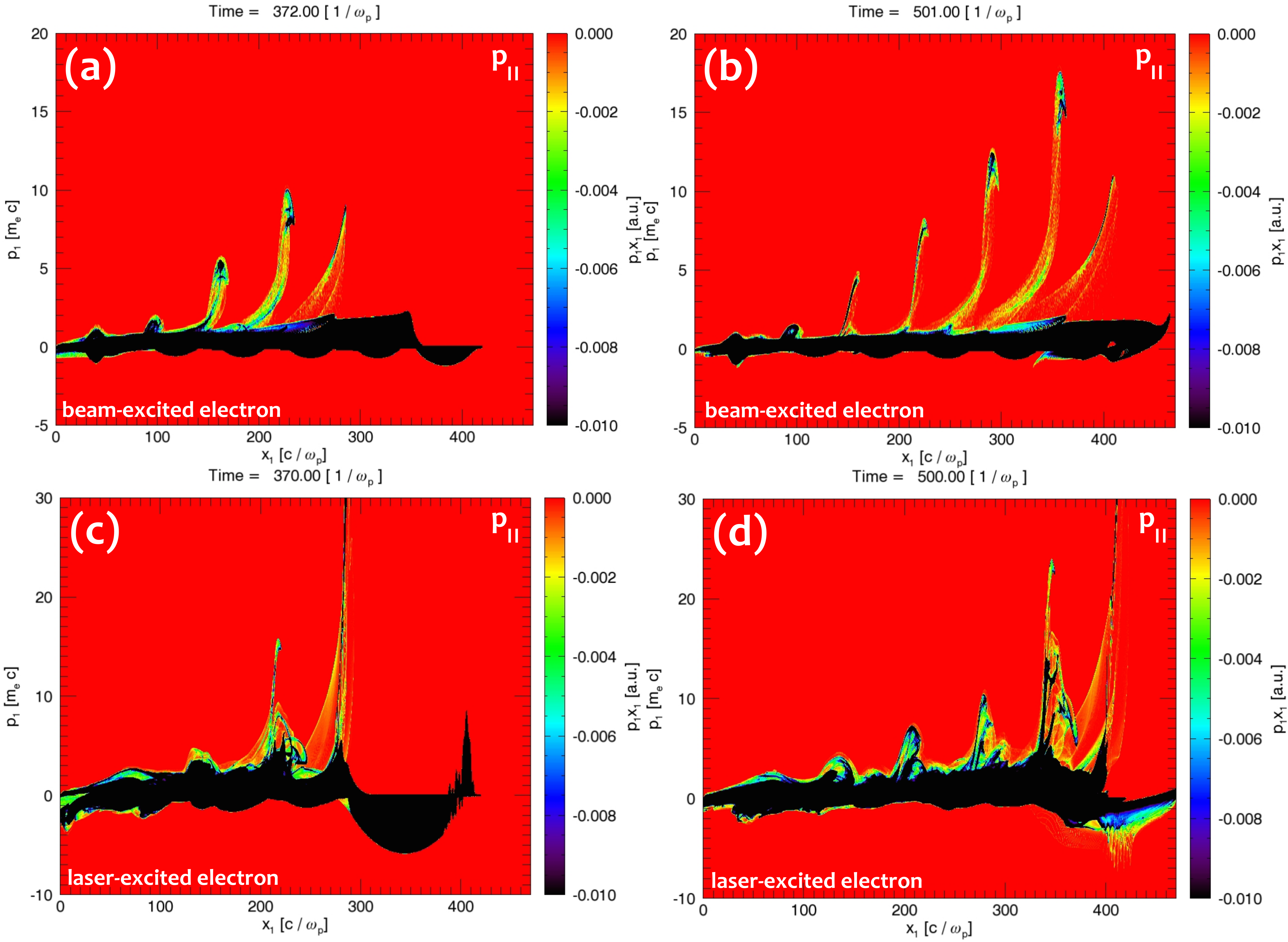}
	\end{center}
\caption{ {\it Time evolution of plasma $e^-$ longitudinal momentum ($p_1x_1$) in beam and laser driven wake-fields}. The plasma $e^-$ longitudinal momentum phase-space at $t\sim370\frac{1}{\omega_{pe}}$ and $t\sim500\frac{1}{\omega_{pe}}$.}
\label{fig:long-momentum-time-evolution}
\end{figure*}

\begin{figure*}
	\begin{center}
   	\includegraphics[width=6.5in]{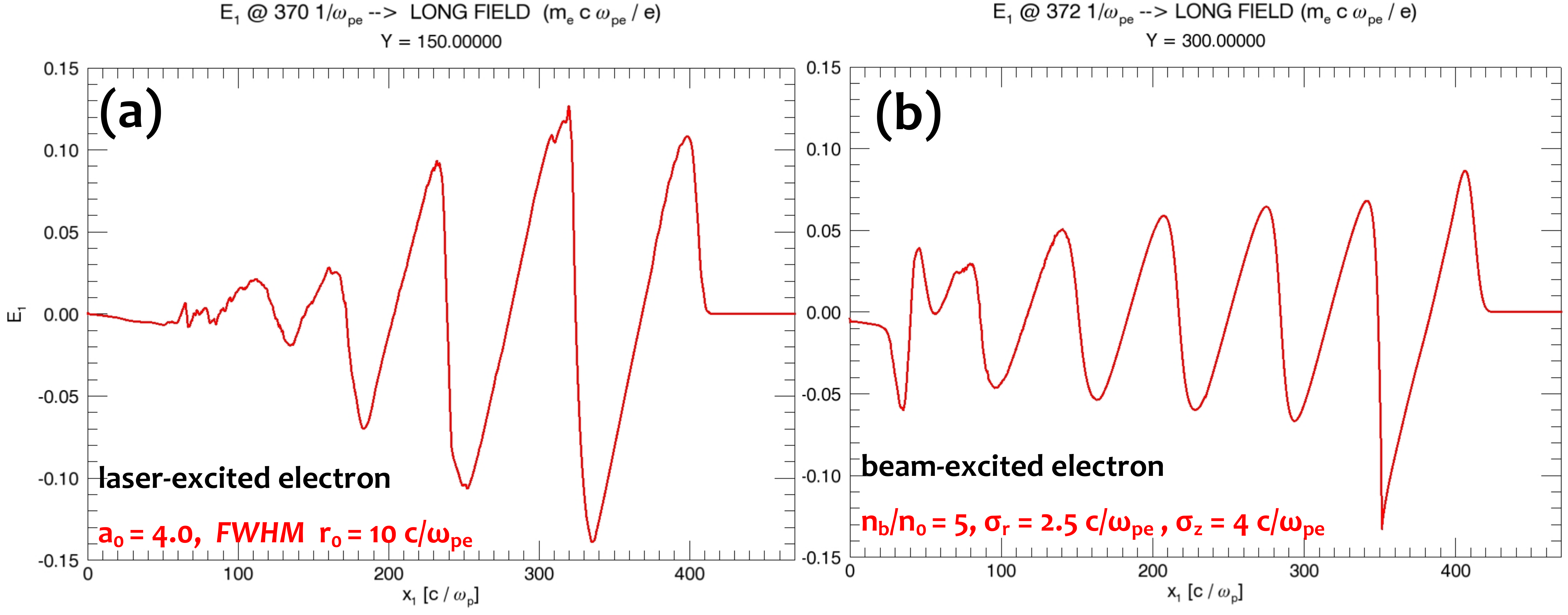}
	\end{center}
\caption{ {\it On-axis longitudinal electric field (e1) for the (a)laser and (b)beam excited plasma}. To compare the laser and beam case the driver parameters are chosen such that the on-axis longitudinal accelerating field is nearly equal. A comparison of the longitudinal fields is shown at 370$\frac{1}{\omega_{pe}}$.}
\label{fig:on-axis-longitudinal-field}
\end{figure*}

To study the trapping of plasma $e^-$ in a rising plasma density gradient at the vacuum-plasma interface driven by an ultra-short laser pulse and a compressed relativistic beam, we use $2\frac{1}{2}D$ OSIRIS\cite{osiris} PIC code with Eulerian specification of the plasma dynamics in a fixed frame. We initialize the homogeneous background plasma density to $n_0 = 0.01n_c$ pre-ionized singly-charged state. The density gradient at the vacuum-plasma interface is in Fig.\ref{fig:vacuum-plasma-gradient}, with vacuum regions in the first and last 50$\frac{c}{\omega_{pe}}$ longitudinal space. We have chosen a relatively high density plasma because of computational convenience as $\omega_{pe}=\omega_0\times\sqrt{n_e/n_c}$. We resolve and reference the real time in simulation to the laser period $2\pi/\omega_0$ thereby the dynamics within a single plasma cycle is simulated in just $\sqrt{n_c/n_e}\sim10$ laser cycles.  We discretize the space with 20 cells per skin-depth ($c/\omega_{pe}$) in the longitudinal and 10 cells per $\frac{c}{\omega_{pe}}$ in the transverse direction. The longitudinal simulation space size is 470 $\frac{c}{\omega_{pe}}$ and the transverse size is 300$\frac{c}{\omega_{pe}}$ for laser-plasma and 600$\frac{c}{\omega_{pe}}$ for beam-plasma simulations. We use absorbing boundary conditions for fields and particles of all species. We use cubic or quartic splines to model the particle shapes. The laser pulse is chosen to be a circularly polarized pulse with normalized vector potential $a_0=4.0$ with a trapezoidal pulse of Gaussian rise and fall time of $10\frac{c}{\omega_{pe}}$ and a flat-time of $20\frac{c}{\omega_{pe}}$. Gaussian-$a(\vec{r})$ with matched\cite{w-lu-prstab-2007} FWHM focal radius of $r_0 = 10\frac{c}{\omega_{pe}}$. The particle beam is initialized with $\gamma\sim38,000$, $\frac{n_b}{n_0}=5.0$ and Gaussian-shape with dimensions of $\sigma_r=2.5\frac{c}{\omega_{pe}}$ and $\sigma_z=4.0\frac{c}{\omega_{pe}}$. These laser and beam driver parameters excite comparable longitudinal fields, in Fig.\ref{fig:on-axis-longitudinal-field}.

\begin{figure*}
	\begin{center}
   	\includegraphics[width=6.5in]{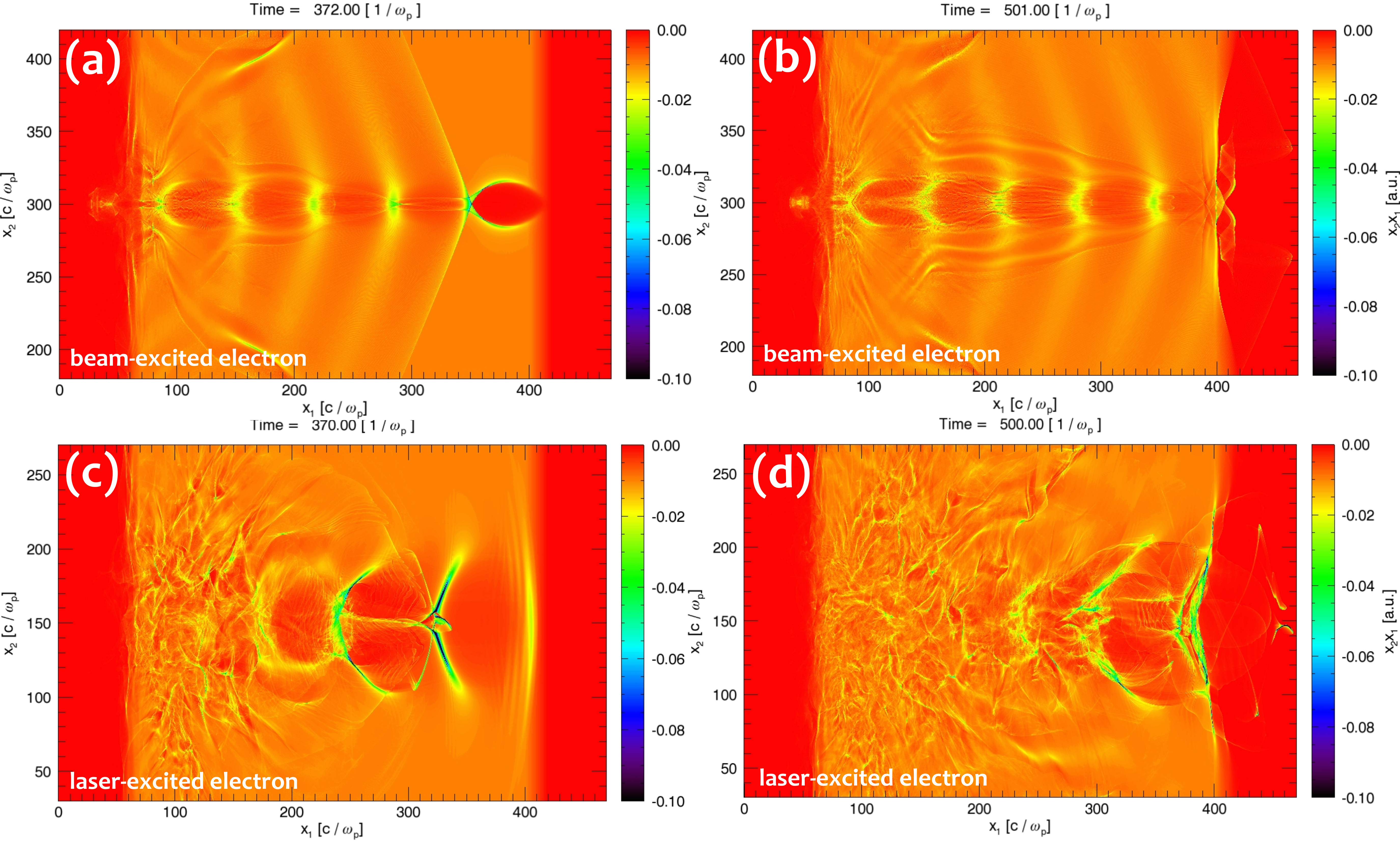}
	\end{center}
\caption{ {\it Self-injected trapped plasma $e^-$ in plasmon-buckets in real-space from 2-D PIC simulations}. The plasma $e^-$ density is shown in real space. In (a) ($t\sim370\frac{1}{\omega_{pe}}$) and (b) ($t\sim500\frac{1}{\omega_{pe}}$) the plasma $e^-$ are trapped only in the second and subsequent buckets in the case of beam-driven plasma. Whereas in corresponding snapshots in (c) and (d), the plasma $e^-$ are trapped in all the buckets in laser-driven case.}
\label{fig:real-space-self-trapping}
\end{figure*}

In Fig.\ref{fig:long-momentum-self-trapping},\ref{fig:long-momentum-time-evolution} the longitudinal momentum phase-space of the plasma $e^-$ is shown. The trapped $e^-$ are seen propagating in longitudinally forward direction, Fig.\ref{fig:long-momentum-time-evolution}. The forward propagating $e^-$ are locked to the peak of the wake-plasmon longitudinal fields, this allows the trapped $e^-$ to continuously gain momentum. In \ref{fig:long-momentum-self-trapping}(a) and (b) the {\it beam-driven} longitudinal momentum of plasma $e^-$ are shown at $t\sim500\frac{1}{\omega_{pe}}$ for longitudinal momentum with longitudinal-dimension ($p_1x_1$) and transverse-dimension ($p_1x_2$) phase-space respectively. Corresponding snapshot of longitudinal momentum phase-space for {\it laser-driven} case are shown in (c) and (d) respectively. It is important to note that the forward longitudinal momentum ($p_1>0$) in the first bucket for the laser with $a_0=4.0$ is much higher than the beam-driven case with $\frac{n_b}{n_0}=5$ (in Fig.\ref{fig:on-axis-longitudinal-field}). The increase (by $\sqrt{\gamma_e}\frac{c}{\omega_{pe}}$) of first-bucket size due to relativistic effects can be observed in Fig.\ref{fig:on-axis-longitudinal-field}(a),\ref{fig:real-space-self-trapping}(c). It is also observed that in the beam-driven case, \ref{fig:long-momentum-self-trapping}(a)-(b), there is no plasma $e^-$ trapping in the first bucket behind the beam. The plasma $e^-$ are trapped in the rising density gradient at the vacuum-plasma interface in the second and subsequent buckets. However, in \ref{fig:long-momentum-self-trapping}(c)-(d), plasma $e^-$ are trapped in all the buckets. In the laser-driven case, the plasma oscillations are non-linear due to relativistic effects and larger ponderomotive effect. So, trapping of plasma $e^-$ occurs in the first bucket due to trajectory distortion (trajectories cross in the back of the bubble) as a result of non-linearity of longitudinal trajectories. However, the trapping in the second bucket is due to the rising density gradient at the vacuum-plasma interface.

In Fig.\ref{fig:real-space-self-trapping} the plasma $e^-$ density is shown in real-space. In \ref{fig:real-space-self-trapping}(a)-(b), the beam-driven plasma is shown in real-space. In laser-drive case in \ref{fig:real-space-self-trapping}(c)-(d), the trapped $e^-$ due to trajectory crossing in the back of the first bucket can be observed. 

\section{Acknowledgement}

Work supported by National Science Foundation, NSF-PHY-0936278 and Department of Energy, DE-SC0010012. We acknowledge the use of 256-node {\it Chanakya} cluster of the Pratt school of engineering at Duke University.


\end{document}